\documentclass[12pt]{article}
\usepackage{epsfig}
\usepackage{amssymb}
\def\beq{\begin{equation}}
\def\eeq#1{\label{#1}\end{equation}}
\def\eeqn{\end{equation}}

\def\beqa{\begin{eqnarray}}
\def\eeqa#1{\label{#1}\end{eqnarray}}
\def\eeqan{\end{eqnarray}}

\let\bar=\overbar

\def\Dslash{\not{\hbox{\kern-4pt $D$}}}
\def\dslash{\not{\hbox{\kern-2pt $\del$}}}

\def\msb{{\bar{\ssstyle M \kern -1pt S}}}

\def\Title#1{\begin{center} {\Large {\bf #1} } \end{center}}

\begin{document}

\Title{\hspace{-.3cm}Obtaining $|V_{ub}|$ exclusively: a theoretical perspective}
\bigskip
\noindent Proceedings of CKM 2012, the 7th International Workshop on the CKM 
Unitarity Triangle, University of Cincinnati, USA, 28 September - 2 October 2012 
\bigskip\bigskip\\
\begin{raggedright}  
{\it Aoife Bharucha\index{Bharucha, A.}\\
II. Institut f\"{u}r Theoretische Physik\\
University of Hamburg\\
Luruper Chaussee 149\\
D-22761 Hamburg, GERMANY}\\
\end{raggedright}
\section{Introduction}
Recent inclusive determinations of $|V_{ub}|$ have uncertainties of approximately
10\%~\cite{VubInc}, as opposed to $\lesssim2\%$ on $|V_{cb}|$
via $B\to X_c l \nu$~\cite{VcbInc}. Exclusive channels provide a competitive 
alternative route to $|V_{ub}|$, but although experimentally more promising this requires information 
about hadronic matrix elements via form factors. Form factors are calculable 
via non-perturbative techniques such as Lattice QCD (see e.g. refs.~\cite{LQCD}) or QCD
sum rules on the light-cone (LCSR). Predictions are usually confined 
to a particular region of $q^2$, the momentum transfer squared, i.e. LCSR and Lattice are restricted 
to large and small recoil energies of the daughter hadron respectively.
In LCSR one considers a correlator $\Pi_\mu$ of the time-ordered product of two quark currents, sandwiched 
between the final state hadron, which is on shell, and the vacuum~\cite{Balitsky:1989ry}, i.e. for a $B$ 
decaying to a $\pi$ of momenta $p_B$ and $p$,
\begin{equation}
\label{eq:correlator} \Pi_\mu=i\,m_b\int d^D x e^{-i\,p_B\cdot x} \langle\pi(p)| T \{\bar{u}(0)\gamma_\mu b(0)\bar{b}(x)i \gamma_5d(x)\}|0\rangle.\\
\end{equation}
This can be expressed on one hand by a light-cone expansion
via perturbative hard scattering kernels convoluted with non-perturbative light-cone distribution 
amplitudes (LCDAs), ordered in increasing twist, or by inserting a sum over excited states, 
i.e. the $b$ hadron and a continuum of heavier states. 
Assuming quark hadron duality above a certain continuum threshold $s_0$, one can 
subtract this continuum contribution from both sides. Borel transforming this 
relation then ensures that this assumption, and the truncation of the series, 
have a minimal effect on the resulting sum rule.
At present, $|V_{ub,excl}|$ is obtained most precisely from 
$B\to\pi l\nu$, where in the limit of massless leptons the decay rate for $B\to\pi$ depends on
a single form factor $f_+(q^2)$. However by considering other channels, e.g. baryonic decays 
such as $\Lambda_b\to p l\nu$, one can obtain interesting 
complementary information.\footnote{In the limit of massless leptons the decay rate for 
$\Lambda_b\to p l\nu$ depends on four form factors, $f_{1,2}(q^2)$ and $g_{1,2}(q^2)$.}
Here I will discuss recent progress in the calculation of the form 
factors for $B\to\pi l\nu$~\cite{Bharucha} and $\Lambda_b\to p l\nu$~\cite{Khodjamirian:2011ub} 
using LCSR.

\section{Recent LCSR updates on $f_+(q^2)$ for $B\to\pi l\nu$}

There has been much progress in the LCSR calculations of $f_+(q^2)$ in the last
15 years.
The next-to-leading order (NLO) corrections to $f_+(q^2)$ at leading twist (twist-2) were first
calculated in LCSR in ref.~\cite{BBB98} and LO corrections up to
twist-4 were calculated in ref.~\cite{Khodjamirian:2000ds}. Since the LO twist-3 contribution
was found to be large, it was confirmed that the NLO
corrections are under control, using both the pole and $\overline{\rm MS}$ mass for $m_b$~\cite{Tw3NLO}.
In ref.~\cite{Khodjamirian:2011ub}, different values for the 
moments of the twist-2 LCDA were employed, extracted from latest experimental data for 
$F_\pi$ using LCSR. The normalised decay rate integrated over a given range in $q^2$,
\begin{equation}
  \Delta
\zeta(0,q_{\mathrm{max}}^2)=\frac{1}{|V_{ub}|^2}\int_0^{q_{\mathrm{max}}^2}dq^2
\frac{d\Gamma}{dq^2}(\Lambda_b\to p l\nu),
\end{equation}
was then predicted to be $\Delta \zeta(0,12
\mathrm{GeV}^2)=4.59^{+1.00}_{-0.85}\,\mathrm{ps}^{-1}$, which can be combined with experimental predictions, 
allowing the extraction of $|V_{ub}|$.

Two-loop corrections to the form factor $f_+(q^2)$ at twist-2 were recently calculated in ref.~\cite{Bharucha}. 
In light of the large two-loop sum rules corrections to $f_B$ calculated in ref.~\cite{Jamin}, one
aim of this work was to test the argument that, in obtaining $f_+(q^2)$ via LCSR, radiative corrections 
to $f_+ f_B$ and $f_B$ should cancel when both calculated in sum rules. Due to the technical challenges posed by a full calculation, a subset of 
two-loop radiative corrections for twist-2 contribution to $f_+(0)$ proportional to $\beta_0$ was considered, 
as this gauge invariant subset is thought to be a good approximation to the complete
next-to-next-to-leading order (NNLO) result. In combination with the experimental result for $f_+(0)|V_{ub}|$ 
one can then obtain $|V_{ub}|$. The necessary diagrams are obtained by inserting a fermion bubble in 
the gluon propagator of the NLO twist-2 diagrams, further details can be found in ref.~\cite{Bharucha}.
% \begin{figure}[t]
% \begin{center}
%  \includegraphics[width=0.55\textwidth]{fig2}
% \end{center}
% \caption{$f_B(0)$ at $\mathcal{O}(\alpha_s^2\beta_0)$ as a function of the Borel
% parameter $M^2$, for central values of input parameters (solid) with
% uncertainties (dotted) calculated as described in the text for $f_+(0)$. This is
% compared to the $\mathcal{O}(\alpha_s)$ result calculated using $s_0=34.2
% \,\mathrm{GeV}^2$ (dashed).}\label{fig:2}
% \end{figure}
\begin{figure}
 \centering
  \includegraphics[width=0.52\textwidth]{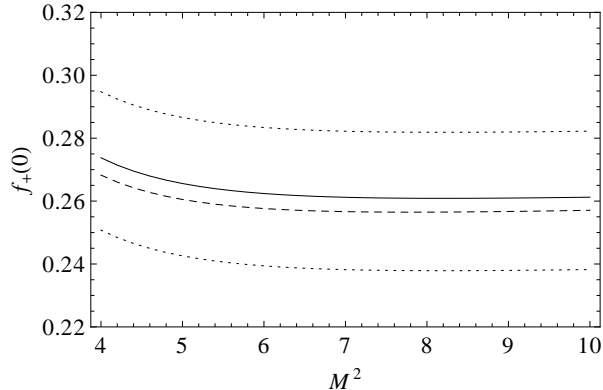}
 \caption{$f_+(0)$ at $\mathcal{O}(\alpha_s^2\beta_0)$ for central values of
input parameters (solid) with uncertainties (dotted), compared to the
$\mathcal{O}(\alpha_s)$ result calculated using $s_0=34.3 \mathrm{GeV}^2$
(dashed), as a function of the Borel parameter $M^2$.}\label{fig:3}
\end{figure}
The results for $f_+(0)$, seen in fig.~\ref{fig:3}, show that despite the 
$\sim 9\%$ positive NNLO corrections to the QCD sum rules result for $f_B$, 
the LCSR prediction for $f_+(0)$ is stable, increasing by $\sim 2\%$ to 
$f_+(0)=0.261^{+0.020}_{-0.023}$, as shown in fig.~\ref{fig:3}. 
This enforces the stability of LCSR with respect to higher order corrections,
and could be taken to provide confirmation that $f_B$ from sum rules, not 
Lattice should be used here.
A recent analysis by BaBar~\cite{Babar2012} finds $|V_{ub}|=(3.34\pm0.10\pm0.05+^{+0.29}_{-0.26})
10^{-3}$ using this result, and $|V_{ub}|=(3.46\pm 0.06\pm
0.08^{+0.37}_{-0.32})10^{-3}$ using $\Delta \zeta(0,12\,\mathrm{GeV}^2)$ from ref.~\cite{Khodjamirian:2011ub}, 
which are clearly in good agreement.

\section{Improvements on form factors for $\Lambda_b\to p$ decays}
Recently there has been increasing work on extracting $|V_{ub}|$ via $\Lambda_b\to p l \nu$.
A number of complications arise in LCSR when baryons are considered instead of mesons, 
the first being the choice of the heavy-light baryon interpolating current $\eta$ described by $\Gamma_b$
and $\tilde{\Gamma}_b$,
\begin{equation}
 \eta=\epsilon^{ijk}(u_iC\Gamma_bd_j)\tilde{\Gamma}_bc_k,
\end{equation}
debated since the 1980s. Additionally, the contribution of the negative parity $\Lambda_b^*$ baryon, 
with $J^P=1/2^-$, which has a similar mass to $\Lambda_b$ is difficult to 
isolate, and in the literature was often included in the continuum~\cite{Yang}.
Recently however it was found to be possible to separate the $\Lambda_b^*$ 
from the $\Lambda_b$ contribution in the sum rule, and on 
comparing results for both $\Gamma_b=\gamma_5$($\gamma_5\gamma_\lambda$) and
$\tilde{\Gamma}_b=1$($\gamma_\lambda$), it was found that the resulting form factors show a reduced 
dependence on the choice of $\Gamma_b$ and $\tilde{\Gamma}_b$~\cite{Khodjamirian3}.
\section{Summary and Outlook}
Recent progress on the LCSR calculation of form factors for the exclusive determination of 
$|V_{ub}|$ was presented. This included recent updates on $f_+(q^2)$:
the 2011 NLO analysis in the $\overline{\rm MS}$ scheme resulted in $|V_{ub}|=(3.46\pm 0.06\pm
0.08^{+0.37}_{-0.32})10^{-3}$ and the 2012 $\mathcal{O}(\alpha_s^2 \beta_0)$ result
found a $\sim 2\%$ increase in $f_+(0)=0.262^{+0.020}_{-0.023}$, 
such that $|V_{ub}|=(3.34\pm0.10\pm0.05+^{0.29}_{-0.26})10^{-3}$.
New results for the form factors for $\Lambda_b\to p l\nu$ were also discussed,
where it was showed that by isolating and removing the negative parity baryons' contribution, the form factors 
show a reduced dependence on the choice of $\Gamma_b$ and $\tilde{\Gamma}_b$.
Future work should focus on combining the $\mathcal{O}(\alpha_s^2 \beta_0)$ $f_+(0)$ and Lattice results
 to determine $|V_{ub}|$ and calculating remaining twist-2 NNLO corrections to $f_+(q^2)$ and gluon radiative corrections
 to the $\Lambda_b\to p$ form factors.

\end{document}